\documentclass[a4paper,12pt]{article}
\usepackage{amsmath,amsfonts,amssymb,epsfig,subfigure}

\renewcommand{\le}{\leqslant}
\renewcommand{\ge}{\geqslant}

\newcommand{\be}{\begin{equation}}
\newcommand{\en}{\end{equation}}

\newcommand{\filler}{\hspace*{\fill}}

\newcommand{\ee}{\textrm{e}}
\renewcommand{\vec}[1]{\boldsymbol{#1}}
\begin{document}
\numberwithin{equation}{section}
%++++++++++++++++++++++++++++++++++++++++++++++++++=++++++++++++++++++
\title{Inhomogeneous shear of orthotropic incompressible
non-linearly elastic solids: \\
singular solutions \\and
biomechanical interpretation}
%++++++++++++++++++++++++++++++++++++++++++++++++++++++++++++++++++++

\author{
  M.~Destrade$^{a,*}$,
  G.~Saccomandi$^{b}$,
  I. Sgura$^{c}$ \filler\\
  {\it\small $^a$Institut Jean Le Rond d'Alembert, CNRS (UMR7190),}
         \filler\\[-6pt]
  {\it\small  Universit\'e Pierre et Marie Curie,}
          \filler\\[-6pt]
  {\it\small Case 162, 4 Place Jussieu, 75252 Paris Cedex 05,
       France;}\filler\\
  {\it\small $^b$Dipartimento di Ingegneria Industriale,}
  \filler\\[-6pt]
  {\it\small Universit\`{a} degli Studi di Perugia,
          06125 Perugia, Italy.}\filler\\
{\it\small $^c$Dipartimento di Matematica,}
  \filler\\[-6pt]
  {\it\small Universit\`{a} del Salento, 73100 Lecce, Italy}
         \filler}

\date{}

%++++++++++++++++++++++++++++++++++++++++++++++++++++++++++
\maketitle
%++++++++++++++++++++++++++++++++++++++++++++++++++++++++++

%%%%%%%%%%%%%%%%%%

\bigskip

\begin{abstract}

We present a detailed study of rectilinear shear deformation in the framework of orthotropic 
nonlinear elasticity, under Dirichlet and mixed-boundary conditions. 
We take a slab made of a soft matrix, reinforced with two families of extensible fibers. 
We consider the case where the shear occurs along the bissectrix of the angle between 
the two privileged directions aligned with the fibers. 
We show that if the two families of parallel fibers are mechanically equivalent,
then only smooth solutions are possible, whereas if the mechanical differences among the 
two families of fibers is pronounced, then strain singularities may develop. 
We determine the precise conditions for the existence of such singular solutions for the standard 
reinforcing orthotropic model. 
We then extend our findings to some orthotropic models of interest in biomechanical applications,
and we discuss the possible relevance of the singular solutions to biomechanics. 

\end{abstract}

\newpage

%%%%%%%%%%%%%%%%%

%%%%%%%%%%%%%%%%%%%%%%%

\section{Introduction}

%%%%%%%%%%%%%%%%%%%%%%%

Biological soft tissues exhibit complex mechanical behaviors, which
are not easily accounted for by classic elastomeric constitutive models.
The extension of the  mathematical models of nonlinear elasticity from
rubber to soft tissues continues to be a challenging area in
theoretical biomechanics.
For example, the presence of oriented collagen
fiber bundles in blood vessels calls for the consideration
of \emph{anisotropy} in the mathematical modeling of the mechanics of
arterial tissues, but the mathematical theory of nonlinear hyperelastic
anisotropic materials is not as developed as the theory of isotropic
nonlinear elasticity.
A consultation of the eminent book by Antman \cite{Antm95} shows that, in
recent years, there has been very few additions to the classical works
of Adkins \cite{Adki55} or of Ericksen and Rivlin \cite{ErRi54} with respect to the
solution of boundary-value problems in nonlinear anisotropic
elasticity.

It is well known that certain radial anisotropies in linear and
non-linear elasticity problems can give rise to stress
singularities which are absent in the corresponding isotropic
version of these problems.
Lekhnitskii \cite{Lekh68} was perhaps the first to
observe this peculiarity, by studying a circular orthotropic plate
compressed by a uniformly distributed force along its external
boundary.
Antman and coworkers (see for example \cite{AnNe87})
extended in some sense this analysis to radially symmetric
equilibrium states of anisotropic nonlinearly elastic bodies.
Another example of this extension to nonlinear elasticity is found
in the paper by Kassianadis \emph{et al.} \cite{KMOP07} on the
finite azimuthal shear of transversely isotropic materials.

Merodio \emph{et al.} \cite{MeSS07} investigated a simple model
for a \emph{nonlinear, transversely isotropic, elastic solid} and
discovered a new kind of singular behavior, not present
in isotropic materials.
It occurs for the \emph{inhomogeneous rectilinear shear} of an
incompressible elastic slab reinforced by a family of
parallel fibers.
They show that, depending on the reinforcement
strength and on the fiber orientation with respect to the shearing
direction, weak solutions for this simple boundary value problem
may be expected.
These solutions are associated with fiber kinking
and loss of ellipticity of the field equations. The deformation
field is continuous, but it suffers a jump in the first derivative
and a blow-up for the second derivative. Therefore the stress
field suffers a discontinuity of first kind, a
phenomenon clearly associated with \emph{mechanical instabilities}.
It also puts into question the applicability of finite element methods
to nonlinear anisotropic elasticity, because the obtention of numerical
solutions to the governing equations requires the calculation of
second-, fourth-, and sometimes higher-order derivatives.
In  biomechanical applications of the constitutive
models of arterial walls, the appearance of stress singularities is
an important mathematical aspect of the theory, because it may be
associated with some pathological states of the tissues
(such as the bursting of an aneurysm).

The aim of the present paper is to
extend the results of Merodio \emph{et al.}  \cite{MeSS07} from
\emph{transverse isotropy} (one family of parallel fibers) to
\emph{orthotropy} (two families of parallel fibers), often encountered
in biological soft tissues.
We note that Fosdick and Royer-Carfagni  \cite{FoRo01} show
that Lekhnitskii's classical solution predicts the interpenetration of
material regions, an unacceptable deformation
behavior in the classical theory of elasticity.
However the solutions proposed in  \cite{MeSS07}
and here are \emph{isochoric} and thus satisfy the local
injectivity requirement.

The paper is organized as follows.
In the next section we write down the governing equations and boundary conditions, and we discuss
the basic mathematical issues at play.
Section 3 is devoted to one of the simplest model of
nonlinear orthotropic elastic materials
(the \emph{standard reinforcing model}), obtained by
adding the classical neo-Hookean strain energy density to two terms that
take into account the reinforcements along the fiber directions.
These latter terms are quadratic in the squared extension along the
fibers.
We solve the problem of inhomogeneous rectilinear shear along the bissectrix to the fibers, 
first for Dirichlet boundary conditions and next for mixed boundary conditions. 
We also provide an energy analysis of the solutions.
In Section 4 we consider a more advanced constitutive
model of the biomechanics literature, 
proposed by Holzapfel \emph{et al}  \cite{HoGO00},
where the reinforcement terms in the strain-energy density
are exponential, in order to account for a strong stiffening
effect (the \emph{artery model}).

The results suggest that orthotropic fiber reinforcement
is quite efficient at cancelling the singularities and the shear
discontinuities encountered in transversally isotropic fiber
reinforcement.
Indeed we recover the main features discovered by
Merodio \emph{et al.} \cite{MeSS07}
(jump in the shear, blow-up of the second derivative of the
displacement)
but under the condition that one family of fibers is much stiffer than
the other.
For the standard reinforcing model, one stiffness modulus
must be at least 9.9 times larger than the other;
for the artery model, the stiffnesses ratio is even higher, due to
exponential terms.
In general, the families of parallel collagen fibers found in arteries
are determined experimentally to be \emph{mechanically equivalent},
suggesting that singularities do not develop, at least in physiological conditions, for the rectilinear shear
of arteries.

%%%%%%%%%%%%%%%%%%%%%%%%%

\section{Basic equations}

%%%%%%%%%%%%%%%%%%%%%%%%%

We consider a composite incompressible slab with thickness $L$, made of
an iso\-tro\-pic matrix reinforced with two families of parallel extensible fibers
(the fibers are all orthogonal to the boundaries of the solid.)
In the undeformed configuration, we call $( X_1, X_2, X_3)$
a set of Cartesian coordinates
such that the solid is located in the $ 0 \le X_3 \le L$ region.
We denote by $\vec{E}_1$,  $\vec{E}_2$,  $\vec{E}_3$ the orthogonal unit
vectors defining the Lagrangian (reference) axes, aligned with the
$X_1$, $X_2$, $X_3$ directions, respectively.

When the solid is sheared in the direction of $\vec{E}_1$,
the particle initially at $\vec{X}$ moves to its
current position $\vec{x}$.
We call $\vec{F} = \partial \vec{x}/\partial \vec{X}$ the associated
deformation gradient tensor, and $\vec{B}= \vec{F}^t \vec{F}$ the left Cauchy-Green
strain tensor.
We then call ($x_1, x_2, x_3$) the Cartesian coordinates, aligned
with ($X_1, X_2, X_3$), corresponding to the current position
$\vec{x}$.
In the current configuration, the basis vectors are $\vec{e}_1$,
$\vec{e}_2$, $\vec{e}_3$, and here they are such that
$\vec{e}_i \equiv \vec{E}_i$ ($i=1,2,3)$.
The deformation is given in all generality by
\be \label{shear_deformation}
x_1 = X_1 +  L f(X_3/L), \qquad x_2 = X_2, \qquad x_3 = X_3,
\en
where $f$ is a yet unknown function of $\eta \equiv X_3/L$ only.
The \emph{amount of shear} is $f' = \text{d}f/\text{d}\eta$.
The deformation \eqref{shear_deformation} is a \emph{simple shear}
when $f'$ is a constant; otherwise it is a \emph{rectilinear
inhomogeneous shear}.
The \emph{direction of shear} is that of $\vec{e}_1 = \vec{E}_1$ and
the \emph{plane of shear} is that of
$(\vec{e}_1 = \vec{E}_1, \vec{e}_2 = \vec{E}_2)$.

We find in turn that
\be \label{B}
\vec{F} = \vec{I} + f' \vec{e}_1 \otimes \vec{E}_3,
\qquad
 \vec{B} = \vec{I} + f'(\vec{e}_1 \otimes \vec{e}_3
  + \vec{e}_1 \otimes \vec{e}_3)  + (f')^2 \vec{e}_1 \otimes \vec{e}_1.
\en
The first principal isotropic strain invariant
$I_1 \equiv \text{tr }\vec{B}$ is given here by
\be
I_1 = 3 + (f')^2,
\en
and the second principal isotropic strain invariant,
$I_2 \equiv [I_1^2 - \text{tr }(\vec{B}^2)]/2$, is also equal to
$3 + (f')^2$.

We call  $\Phi$ ($\Psi$, respectively) the angle between the
direction of one family of parallel fibers (the other family,
respectively) and the direction of shear $X_1$.
In other words, the unit vectors $\vec{M}$ and $\vec{N}$ (say)
in the two preferred fiber directions have components
\be
\vec{M} = \cos \Phi \vec{E}_1 + \sin \Phi \vec{E}_3, \qquad
\vec{N} = \cos \Psi \vec{E}_1 + \sin \Psi \vec{E}_3,
\en
and they are transformed into $\vec{m} = \vec{F M}$ and $\vec{n} =
\vec{F n}$ in the current configuration,
\be
\vec{m} = (\cos \Phi + f' \sin \Phi)\vec{e}_1 + \sin \Phi \vec{e}_3,
\qquad
\vec{n} = (\cos \Psi + f' \sin \Psi)\vec{e}_1 + \sin \Phi \vec{e}_3.
\en

In the remainder of the paper, we restrict our
attention to the special case where the material is
\emph{sheared along a bisectrix
of the angle between the two families}.
Generality is lost with this approach, but it has the merit of
keeping low the number of geometric parameters;
we also argue that it still captures some salient
features of sheared soft tissues with two preferred directions.

\begin{figure}
 \centering
\includegraphics[width=0.9\textwidth]{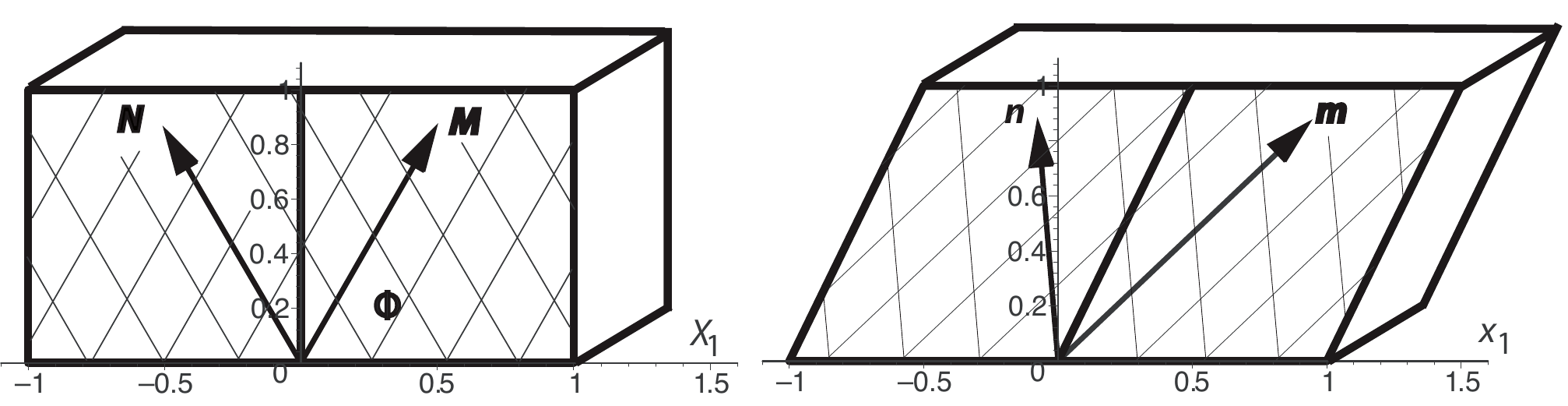}
 \caption{{\small Two unit squares lying in the transverse section
 of a slab reinforced
 with two families of fibers (thin lines) and subject to a simple shear
of amount $0.5$
 along the bissectrix of the angle between the two families.
 In the reference configuration, one family of fibers is aligned with
the unit vector $\vec{M}$
 making an angle $\Phi = 60^\circ$ with the $X_1$-axis; the other family
is aligned with
 $\vec{N}$, at an angle $120^\circ$
 with the $X_1$-axis.
 In the current configuration, they are along $\vec{m}$ and $\vec{n}$,
respectively.}}
 \label{fig_two_families}
\end{figure}

Hence from now on, $\Psi = \pi  - \Phi$ and
the angle between the two preferred directions is $\pi - 2 \Phi$.
In other words, the unit vectors $\vec{M}$ and $\vec{N}$
in the preferred fiber directions have components
\be \label{M_N}
\vec{M} = \cos \Phi \vec{E}_1 + \sin \Phi \vec{E}_3,
\qquad
\vec{N} = -\cos \Phi \vec{E}_1 + \sin \Phi \vec{E}_3,
\en
in the reference configuration, and they are transformed
into
\be \label{m_n}
\vec{m} = (\cos \Phi + f' \sin \Phi)\vec{e}_1 + \sin \Phi \vec{e}_3, \qquad
 \vec{n} = (-\cos \Phi + f' \sin \Phi)\vec{e}_1 + \sin \Phi \vec{e}_3,
\en
in the current configuration;
Figure \ref{fig_two_families} is a visualization of the situation in the
case of a simple (homogeneous) shear of amount $0.5$ and
an angle $\Phi = 60^\circ$.
Note that because the reinforcements are not directional, we may without
loss of generality restrict ourselves to the range $0 < \Phi < \pi$.

We now introduce the \emph{anisotropic invariants} $I_4 \equiv \vec{m
\cdot m}$
and $I_5 \equiv \vec{F m \cdot}\vec{F m}$;
in particular we find
\be \label{I4}
 I_4 = 1 + f' \sin 2 \Phi + (f')^2 \sin^2 \Phi.
\en
Recall that $I_4$ is the squared stretch in the fiber direction
(Spencer  \cite{Spen72}).
In particular, if $I_4 \ge 1$ then the fibers aligned with $\vec{m}$ are
in extension,
and if $I_4 \le 1$ then they are in compression.
Clearly here, when $0 \le \Phi \le \pi/2$, the quantity
$I_4 -1$ is always positive and the fibers  aligned with $\vec{m}$ are
in extension.
On the other hand, when $\pi/2 < \Phi < \pi$, there always
exist a certain amount of shear (explicitly, $-2 /\tan \Phi$)
above which these fibers are in compression.

The other anisotropic invariants are
$I_6 \equiv \vec{n \cdot n}$,
$I_7 \equiv \vec{F n \cdot}\vec{F n}$,
and $I_8 \equiv \vec{m \cdot n}$.
Here we find that
\be
 I_6 = 1 - f' \sin 2 \Phi + (f')^2 \sin^2 \Phi.
\en

In general, the strain-energy density $W$ of a hyperelastic
incompressible
solid reinforced with one or two families of parallel extensible fibers
depends on
two isotropic deformation invariants: $I_1$ and $I_2$, and on the five
anisotropic deformation
invariants  \cite{Spen72, Boel87}: $I_4$, \ldots, $I_8$.
Henceforth we make the assumption that $W$ is the sum of
an isotropic part and an anisotropic part.
For the isotropic part, modelling the properties of the elastin matrix,
we take the neo-Hookean strain-energy density, with constant shear
modulus $\mu$.
For the anisotropic part, modelling the properties of the extensible
collagen
fibers, we take the sum of a function of $I_4$ only
and a function of $I_6$ only, say $F(I_4) + G(I_6)$.
Hence we restrict our attention to those solids with strain energy
density
\be \label{W}
W = \mu(I_1 - 3)/2 + F(I_4) + G(I_6).
\en

Now the Cauchy stress tensor $\vec{\sigma}$ derived from this strain
energy function is (see e.g. Ogden, 1984),
\be
\vec{\sigma} =  - p \vec{I} + \mu \vec{B}
 + 2 F'(I_4) \vec{m} \otimes \vec{m}
 + 2 G'(I_6) \vec{n} \otimes \vec{n},
\en
where $p$ is a Lagrange multiplier introduced by the constraint
of incompressibility, and
$F' \equiv \text{d}F / \text{d}I_4$,
$G' \equiv \text{d}G / \text{d}I_4$.

Because shear is a plane strain deformation, and because the
fibers lie in the plane of shear, it is a simple matter to find the
directions of principal
stresses. One is normal to the plane of shear, and the two others are in
the
($\vec{e}_1, \vec{e}_3$) plane, at the angles $\varphi$ and $\varphi +
\pi/2$ from
the direction of shear, where $\varphi \in ]0, \pi/4]$ is defined by
\be
\tan 2 \varphi = 2(\vec{e}_1 \cdot \vec{\sigma e}_3) /
  (\vec{e}_1 \cdot \vec{\sigma e}_1 - \vec{e}_3 \cdot \vec{\sigma e}_3).
\en
Here we find that
\begin{align}
 & \vec{e}_1 \cdot \vec{\sigma e}_3 = \mu f' + 2 F'(I_4) m_1 m_3 +
     2 G'(I_6) n_1 n_3,
\end{align}
where $m_i \equiv \vec{m \cdot e}_i$ and $n_i \equiv \vec{n \cdot e}_i$
are found from \eqref{m_n}.

The equilibrium equations, $\text{div }\vec{\sigma} = \vec{0}$
(in the abscence of body forces) reduce to
\begin{align}
& \dfrac{\partial p}{\partial x_1} = \dfrac{\text{d}}{\text{d}x_3}
 \left[ \mu f' + 2F'(I_4) m_1 m_3 + 2 G'(I_6) n_1 n_3 \right],
 \\ \notag
& \dfrac{\partial p}{\partial x_2} = 0,
 \\ \notag
& \dfrac{\partial p}{\partial x_3} = \dfrac{\text{d}}{\text{d}x_3}
 \left[ \mu f' + 2F'(I_4) m_3^2 + 2 G'(I_6) n_3^2 \right],
 \end{align}
where the expressions in brackets are independent of $x_1$ and $x_2$.
It follows that
\be
p = p(x_1, x_2) =
 C_0 x_1 + 2F'(I_4) m_3^2 + 2 G'(I_6) n_3^2 + D,
\en
where $C_0$, $D$ are arbitrary constants of integration,
is a suitable pressure field.
A single governing equation remains to be solved for the shear
deformation,
namely
\be \label{equation}
\dfrac{\text{d}}{\text{d}\eta}\left[\mu f' + 2 F'(I_4) m_1 m_3 +
     2 G'(I_6) n_1 n_3 \right] = C_0 L.
\en

We consider two specific \emph{boundary value problems} (BVPs).
In the reference configuration, the slab of thickness $L$
in the $X_3$ direction and of infinite dimensions in the other
directions
is bonded to two infinite rigid plates located at $X_3=0$ and $X_3 =
L$.
A constant \emph{pressure gradient} is applied in the $x_1$ direction
and drives the deformation of the slab.
The overall goal of our investigation is to solve
\eqref{equation} subject \emph{(i)} to the \emph{Dirichlet boundary conditions}:
$f(0)=0,\, f(1)=0$, and \emph{(ii)} to the \emph{mixed boundary conditions}:
$f(0)=0, \, f'(1)= K_1$, where $K_1$ is a prescribed constant.
In Case \emph{(i)}, we have a classical two-point boundary value problem which,
for an isotropic medium, may be reduced
to a Cauchy problem by using symmetry considerations.
In our anisotropic case it may happen that, once the second-order
differential equation
\eqref{equation} is rewritten in normal form, the corresponding right
handside
is neither continuous nor Lipschitzian with respect to $f'$.
Then standard methods  for the study of the existence and
uniqueness of the solution may not apply any longer;
moreover, the solution may develop singularities.
In Case \emph{(ii)}, the BVP is simpler to solve, 
but it is also possible to have non-smooth solutions.
We point out that enforcing the boundary
condition $f'(1) = K_1$ is equivalent to prescribing the shear stress $T_{12}$
on the upper face of the slab.

For transversally isotropic materials, Case \emph{(i)} has been
studied by Merodio \emph{et al} \cite{MeSS07} and a mixed-BVP similar to Case \emph{(ii)} 
has been considered for azimuthal shear by Kassianadis \emph{et al} \cite{KMOP07}.

%%%%%%%%%%%%%%%%%%%%%

\section{The standard reinforcing model}

%%%%%%%%%%%%%%%%%%%%%

%***************************************

\subsection{Normal form of the BVP}

%***************************************

The \emph{standard reinforcing model} for solids with
two family of fibers is a special case of \eqref{W}.
Its strain energy density is
\begin{equation} \label{standard}
W=\mu(I_1-3)/2+\mu E_1(I_4-1)^2/4+\mu E_2(I_6-1)^2/4,
\end{equation}
where $\mu E_1$ and $\mu E_2$ are the extensional moduli in the fiber
directions.
The BVPs based on (\ref{equation})  are now
\begin{equation} \label{equationstandard}
\dfrac{\text{d}}{\text{d} \eta}\left[ f' + E_1(I_4-1) m_1 m_3 +
     E_2(I_6-1) n_1 n_3 \right]= C_0 L / \mu;
\end{equation}
with the boundary conditions \emph{(i)}: $f(0)=0, \, f(1)=1$ and 
\emph{(ii)}: $f(0)=0, \, f'(1) = K_1$.
We begin our study with the Dirichlet boundary conditions, Case \emph{(i)}.

The differential equation may be rewritten as
\begin{equation} \label{equationstandard2}
\dfrac{\text{d}}{\text{d} \eta}\left\{ f' + \gamma f' \sin^2\Phi
[2\cos^2\Phi + 3 \beta (f') \sin \Phi \cos \Phi +
(f')^2\sin^2\phi]\right\}= C_0 L / \mu,
\end{equation}
where we introduced the dimensionless material constants
$\gamma$ and $\beta$, defined as
\be \label{alpha_beta}
\gamma = E_1+E_2, \qquad \beta=(E_1-E_2)/(E_1+E_2).
\en
The quantity $\gamma$ gives a measure of the collagen/elastin strength
ratio, and the quantity $\beta$ gives a measure of the orthotropy.
If $\gamma=0$, then the material is isotropic.
If $\beta = \pm 1$, then either $E_1 = 0$ or $E_2 = 0$ and the
solid is transversally isotropic (there is only one active family of
parallel fibers);
if $\beta = 0$, then $E_1 = E_2$ and the two families of fibers are said
to
be \emph{mechanically equivalent}.

In its normal form, the BVP Case \emph{(i)} reads
\be \label{normalformstand}
\dfrac{\text{d}^2 f}{\text{d} \eta^2} = \dfrac{\alpha}{D(f', \Phi)},
\qquad
f(0)= f(1)=0,
\en
where $\alpha \equiv C_0 L / \mu$ is a dimensionless measure of the
pressure gradient and where
the denominator $D$ is defined as
\be \label{D}
D(f', \Phi) = 1 + \gamma \sin^2\Phi [ 2\cos^2 \Phi + 6 \beta \cos \Phi
\sin \Phi (f')
 + 3\sin^2 \Phi (f')^2].
\en

First we note that when $\beta = 1$, the whole analysis is consistent
with that of
Merodio et al.  \cite{MeSS07}
for a transversally isotropic slab.
Also, when $\beta = 0$, the governing equation coincides with that
obtained for the rectilinear
inhomogeneous shear of the \emph{isotropic} solid slab with strain
energy density
$W=(\mu + 2\gamma \sin^2 \Phi \cos^2\Phi)(I_1-3)/2 + (\gamma \sin^4
\Phi)(I_1-3)^2/4$.
It follows that when the two families of fibers are mechanically
equivalent, only smooth solutions
exist and no singularity may develop.

Next we take $\beta \ne 0$ and notice that $D$ is a quadratic in the
amount of  shear $f'$.
If its discriminant is negative, then no singularity may develop.
The denominator $D$ has real roots when
\be
\left( 3 \beta^2 - 2  \right) \gamma \sin^2{2\Phi}-4 \geq 0.
\en
Therefore a \emph{necessary} condition for the appearance of
singularities is that
\be
\beta^2 > 2/3.
\en
Assume that the fibers along $\vec{M}$ are stiffer than those along
$\vec{N}$.
Then $E_1 > E_2$ and this inequality means that $E_1/ E_2 > 5+2\sqrt{6}
\simeq 9.9$.
Hence we are certain that singularities do not develop  when the
fibers along $\vec{M}$
are less than $9.9$ times  stiffer than the fibers along $\vec{N}$.

%**********************************************

\subsection{Orthogonal fibers: $\Phi = \pi/4$}

%**********************************************

\begin{figure}
 \centering
\includegraphics[width=0.9\textwidth, height=0.45\textwidth]{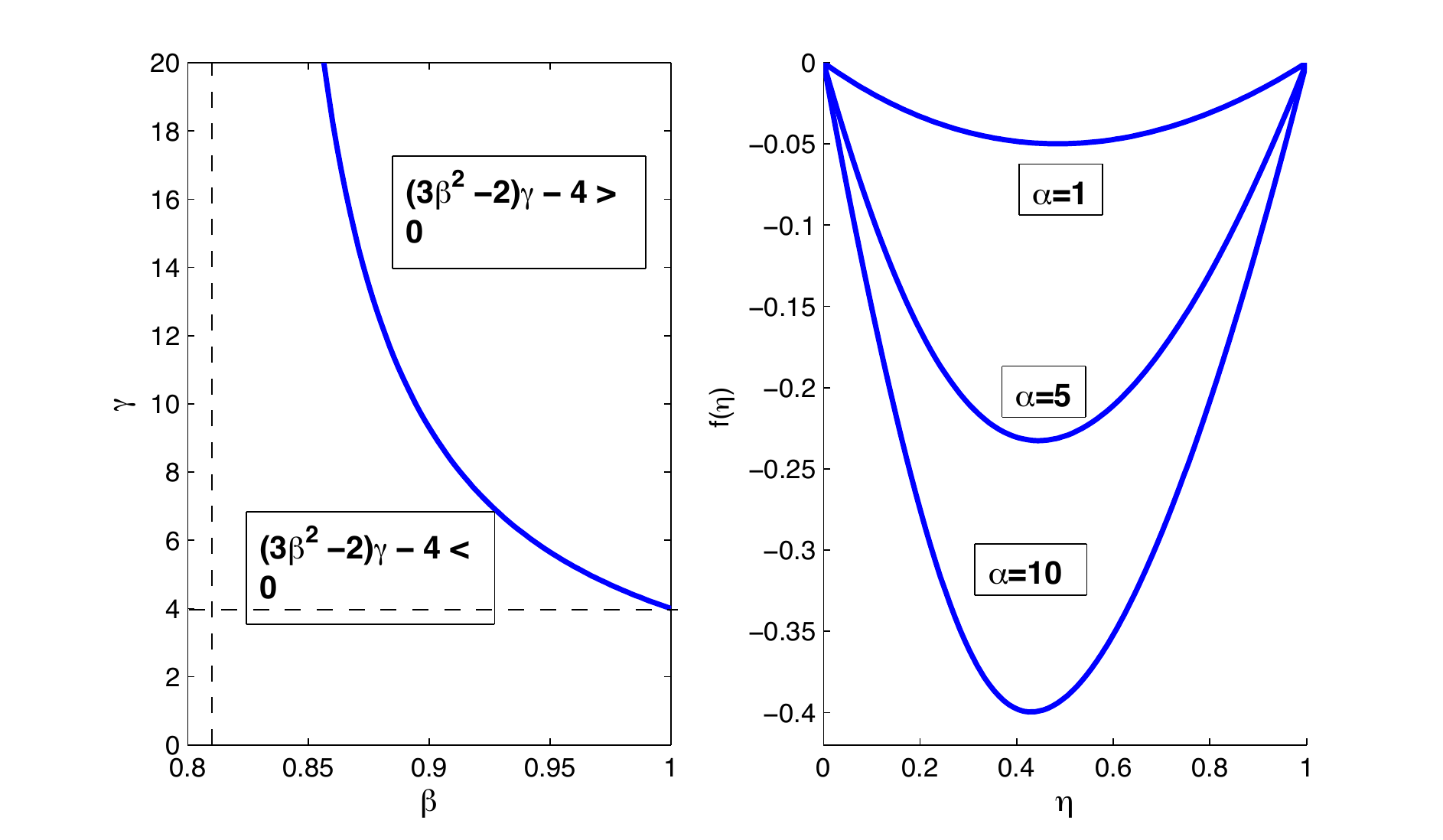}
 \caption{{\small On the left: curve in the  ($\beta, \gamma$) plane separating
the region where only smooth
 exist ($(3 \beta^2 - 2) \gamma - 4 < 0$) from the region where
singularities might develop for the second
 derivative of the displacement, in the case where the two families of
fibers are orthogonal.
 On the right: an example of a completely smooth solution, obtained for
$\beta = 0.5$, $\gamma=3.0$
 and for several values of the pressure gradient, as measured by
$\alpha$.}}
 \label{fig_smooth}
\end{figure}

Here we focus on the special case where one family of fibers is
orthogonal to the other family
($\Phi=\pi/4$).
Then the denominator $D$ in \eqref{D} reduces to
\be \label{quadratic}
D(f', \pi/4) = 1 + (\gamma / 4) \left[2 + 6 \beta (f') + 3(f')^2
\right].
\en
Clearly, whether $f''$ develops singularities or not depends among other
things on the sign
of the quantity $(3 \beta^2 - 2) \gamma - 4$.
Figure \ref{fig_smooth} displays on the left the curve where this
quantity is zero in the
($\beta, \gamma$) plane.
When it is negative, the existence and uniqueness of a smooth solution
are guaranteed by general
theorems and standard numerical procedures of integration can be
implemented.
For instance, we take  $\beta = 0.5$, $\gamma=3.0$, and $\alpha = 1.0,
5.0, 10.0$ in turn, and
obtain the displacements displayed on the right of Figure
\ref{fig_smooth},
using the finite difference method implemented into \textsc{Maple}.

When  $(3 \beta^2 - 2) \gamma - 4 \ge 0$, there is a chance that
singularities may develop within the
thickness of the slab and we now investigate this possibility.
First we consider the case where this discriminant is equal to zero,
when $\gamma = 4/(3\beta^2-2)$.
Then
\be \label{D4}
D = \dfrac{3}{3\beta^2-2}(f' + \beta)^2,
\en
and integrating  \eqref{normalformstand} once gives
\be \label{f'4}
(f' + \beta)^3 = \alpha(3\beta^2 - 2) (\eta - \eta_0) + \beta^3,
\en
where $\eta_0 \in (0,1)$ is a point in the thickness of the slab where
$f' =0$ (its existence is ensured
by the continuity and differentiability of $f$, coupled to the boundary
conditions $f(0)= f(1) = 0$).
Solving for $f'$ gives
\begin{equation} \label{KK}
f'(\eta) = \mp \left[ \beta^3 + \alpha(3 \beta^2-2) (\eta -
\eta_0)\right]^{1/3} - \beta,
\end{equation}
where the sign depends on the sign of the radical.
Integrating further, and imposing $f(0)=0$, we obtain
\begin{multline} \label{ff}
f(\eta) = \dfrac{3}{4 \alpha (3 \beta^2-2)}
  \left\{ \left[\beta^3 + \alpha (3 \beta^2 - 2)(\eta -
\eta_0)\right]^{4/3}
  \right. \\
\left. -\left[\beta^3 - \alpha(3 \beta^2-2) \eta_0 \right]^{4/3}
\right\} - \beta \eta.
\end{multline}
To solve the BVP entirely, it remains to determine $\eta_0 \in (0,1)$.
It is fixed by the second boundary condition: $f(1) = 0$, i.e. it is a
solution to the equation
\be \label{bc2}
  \left[\beta^3 + \alpha (3 \beta^2 - 2)(1 - \eta_0)\right]^{4/3}
 -\left[\beta^3 - \alpha(3 \beta^2-2) \eta_0 \right]^{4/3} =   4 \alpha
\beta (3 \beta^2-2)/3.
\en
Now, collecting \eqref{normalformstand}, \eqref{D4}, \eqref{f'4}, we see
that $f''$ blows up at
$\eta = \eta_S$ given by
\be
\alpha(3\beta^2-2)(\eta_S-\eta_0) + \beta^3=0.
\en
The final condition to impose for this singularity is that it occurs
within the thickness of the slab:
$0 \le \eta_S \le 1$, i.e.
\be \label{ineq4}
0 \le \eta_0 - \dfrac{\beta^3}{\alpha (3\beta^2-2)} \le 1.
\en

 \begin{figure}
 \centering
\includegraphics[width=0.75\textwidth,height=0.45\textwidth]{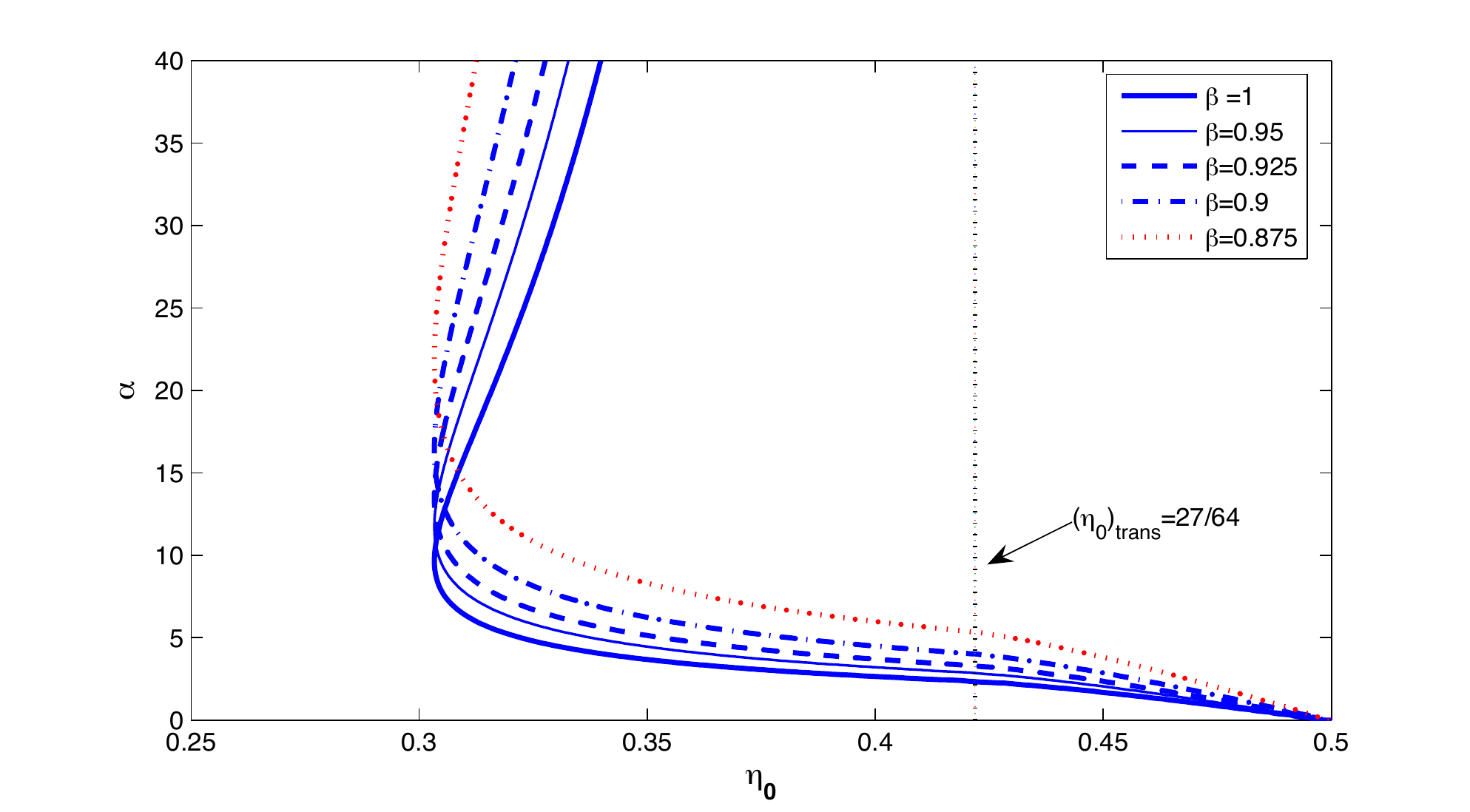}
 \caption{{\small Curves in the ($\eta_0, \alpha$) plane giving the
locus of the constant of integration $\eta_0$
 corresponding to a given level of pressure gradient, as measured by
$\alpha$,
 in the case where the two families of fibers are orthogonal and some
fibers
 are stiff enough to guarantee the appearance of singularities.
 The curves are shown in the first quarter of the plane, for $\beta =
0.875, 0.9, 0.925, 0.95, 1.0$.
 They are antisymmetric with respect to the point $(0.5,0)$ (second part
not shown here).
 The curves are limited to the right by a vertical line at $\eta_0 =
27/64 = 0.421875$.}}
 \label{fig_eta_alpha}
\end{figure}
On Figure \ref{fig_eta_alpha} we graph the curves defining the pairs
($\eta_0, \alpha$)
such that the
second boundary condition \eqref{bc2} is satisfied, for several values
of $\beta$.
We limit the display to the range $\eta_0 < 0.5$
because the curves are antisymmetric with respect to the point
$(0.5,0)$.
For visual reasons, the upper bound is taken as $\alpha_{\text{max}} =
40.0$.
The other limit of each curve is imposed by the inequalities
\eqref{ineq4},
specifically here the lower one.
The corresponding \emph{transitional behavior} is dictated by the
equality
\be
\eta_0 = (\eta_0)_{\text{trans}} \equiv \dfrac{\beta^3}{\alpha(3
\beta^2-2)}.
\en
When this holds, the corresponding transitional level of pressure
gradient is found from \eqref{bc2} as
\be
\alpha = (\alpha)_{\text{trans}} \equiv \dfrac{64 \beta^3}{27(3
\beta^2-2)}.
\en
Substituting back above gives
\be
(\eta_0)_{\text{trans}} = 27/64 = 0.421875.
\en
Hence all the curves stop at the vertical barrier $\eta_0 = 0.421875$,
irrespective of the value of $\beta$.

\begin{figure}
 \centering
\includegraphics[width=0.9\textwidth,height=0.4\textwidth]{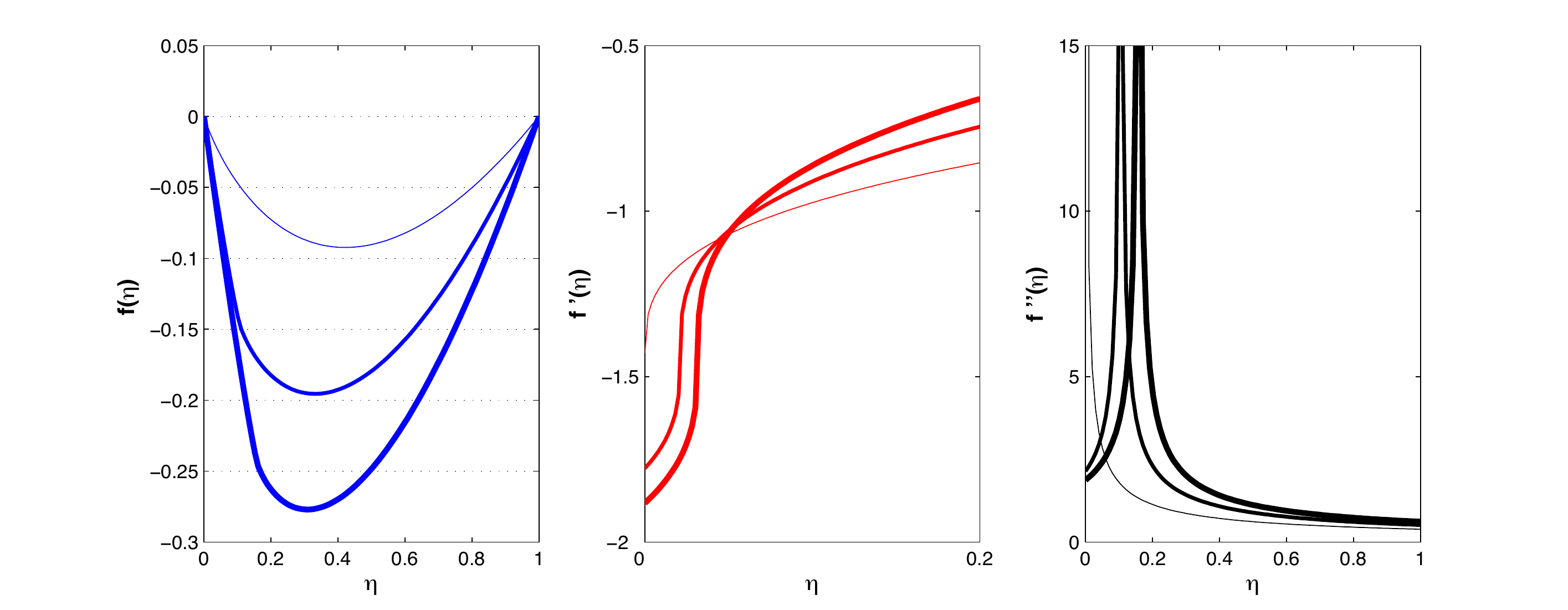}
 \caption{{\small Plots of the displacement and of
 its first and second derivatives
 through the slab thickness in the case where the
 two families of fibers are
 orthogonal and singularities develop.
 Here $\beta=0.875$ (large difference in fiber stiffness)  and $\alpha$
(giving a
 measure of the pressure gradient) and $\eta_0$ (constant of
integration) are
 chosen so that the second derivative is discontinuous.
 Thin curves: $\alpha = 5.3489$, $\eta_0 =  0.421875$);
 Medium thickness curves: $\alpha = 10.0$, $\eta_0 =  0.33120$;
 Thick curves: $\alpha = 15.0$, $\eta_0 =  0.30871$.}}
 \label{fig_4}
\end{figure}

For all values of $\alpha$ and $\eta_0$ such that the point ($\alpha,
\eta_0$) belongs to
one of these curves, a singularity develops within the thickness for the
second derivative of
the displacement.
For instance at the points ( $(\alpha)_{\text{trans}},
(\eta_0)_{\text{trans}}$), the exact solution
\eqref{ff} and its derivatives reduce to
\begin{equation}\label{f_sing_0}
f(\eta) = \beta \left( \eta^{4/3} - \eta \right),
\quad
f'(\eta) = \beta \left[ (4/3) \eta^{1/3}/3 - 1 \right],
\quad
f''(\eta) = (4/9) \beta \eta^{-2/3},
\end{equation}
and $f''$ clearly blows up on the $\eta=0$ face of the slab.
For Figure \ref{fig_4} we take $\beta=0.875$ and
three points on the corresponding  curve of
Figure \ref{fig_eta_alpha} , namely
($\alpha = 5.3489, \eta_0 =  (\eta_0)_{\text{trans}}$),
($\alpha = 10.0, \eta_0 =  0.33120$),
and
($\alpha = 15.0, \eta_0 =  0.30871$).
For the first combination, $f''$ blows up on the
slab face $\eta = 0$;
for the second and third combinations, it blows up
within the thickness of the slab.

Next we consider the case where the discriminant of the $f'$ quadratic
in \eqref{quadratic} is positive:
$(3\beta^2-2)\gamma - 4 >0$, and focus now on finding singularities for
$f'$, the amount of shear.
Integrate the BVP \eqref{normalformstand}, \eqref{quadratic} once to get
\be \label{g}
g(f') = \alpha (\eta - \eta_0),
\en
where $\eta_0$ is a constant of integration, and $g$ is the
following cubic,
\be \label{cubic}
g(x) := \left( 1 + \dfrac{\gamma}{2}\right) x
 + \dfrac{3\beta \gamma}{4} x^2 + \dfrac{\gamma}{4} x^3,
\en
with a local maximum (resp. minimum) at $x_1$ (resp. $x_2$)
defined as
\be \label{x_1}
x_{1,2} = -1 \mp \sqrt{\dfrac{(3\beta^2-2)\gamma-4}{3\gamma}}.
\en
For
$g_1 \equiv g(x_1)$ and $g_2 \equiv g(x_2)$, we find
\be \label{g_1}
g_{1,2} = -\dfrac{1}{6} \left[
2(2+\beta) + \gamma(1-\beta)(3\beta-2)\right]
 \pm \dfrac{\gamma}{2} \left[ \dfrac{(3\beta^2-2)\gamma -4}{3\gamma}
\right]^{3/2}.
\en

\begin{figure}
 \centering
\includegraphics[width=0.9\textwidth,height=0.35\textwidth]{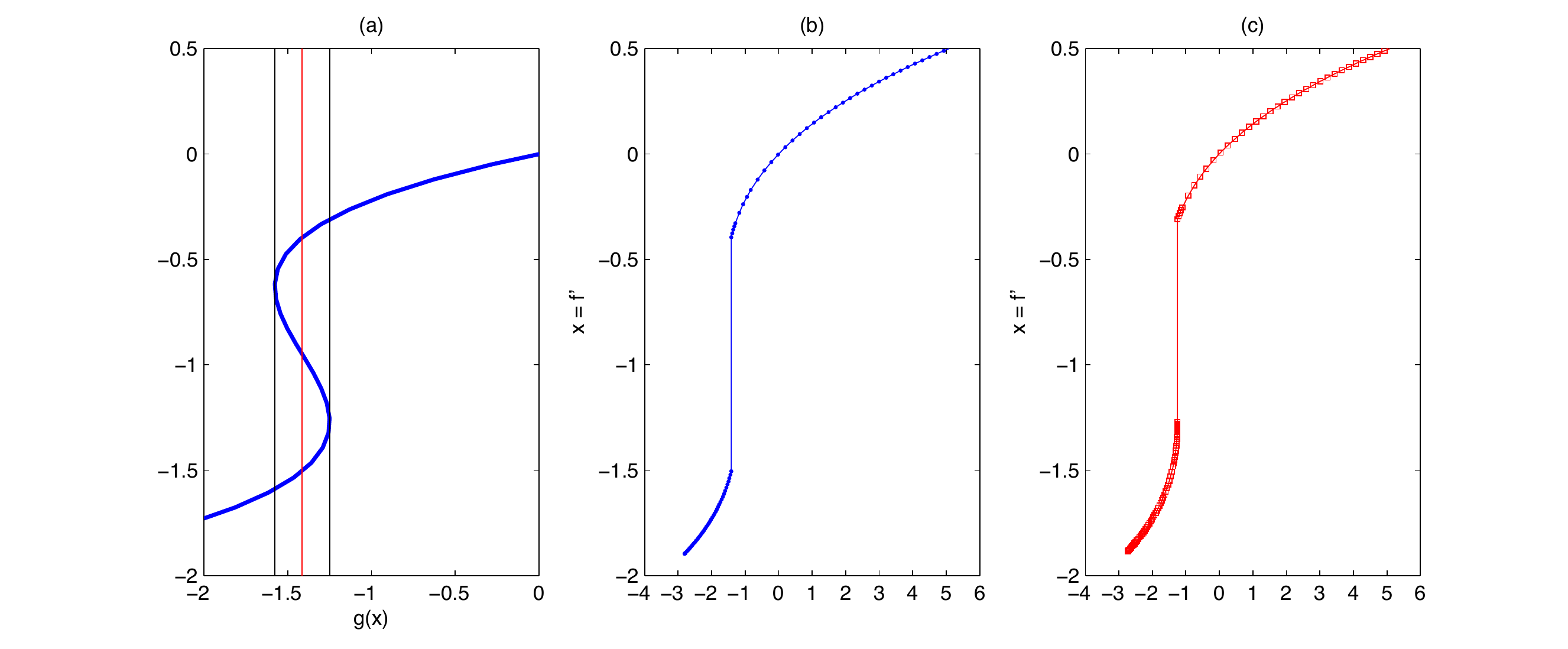}
 \caption{{\small (a) Plot of the cubic establishing a relationship between the amount of shear and 
 the thickness. When $g_1 \le \eta \le g_2$,  there are three possible values of $f'$ for each $\eta$.
 The vertical dashed lines are at $\eta = g_1, g_2$;
 the vertical full line defines two regions of equal area between $g_1$ and $g_2$.
 (b) Maxell rule convention: jump in the amount of shear at the thickness giving equal areas. 
 (c) Maximum delay convention: jump in the amount of shear at $\eta = g_1$.
 Here the Dirichlet BVP is solved for  
 $\beta=0.95$, $\gamma=10.0$, $\alpha = 10.0$.}}
 \label{figWK}
\end{figure}

Clearly, a systematic procedure to establish a one-to-one correspondence between $x$ and $g$
everywhere (or equivalently, between $f'$ and $\eta$) runs into difficulties in the interval 
$x_1 \le x \le x_3$, where $x_3$ is the root of $g(x) = g_1$ other than $x_1$, 
see Figure \ref{figWK}(a).
To address this problem, we take the stance that $f'$ jumps from a low value to 
a higher one. 
In order to jump following the absolute minima of the energy, we consider in turn the 
\emph{Maxwell rule convention} of equal area, see Figure  \ref{figWK}(b), 
and the \emph{Maximum delay convention}, see Figure  \ref{figWK}(c). 
We propose to track these two possible solutions by a suitable numerical approach. 
This hands-on approach is required because usual numerical methods sometimes 
fail in finding good  approximations. 
In fact, commercial code solvers issue a warning here about possible failure in the numerical
convergence and are unable to provide a satisfactory solution in this region.
%This shortcoming is illustrated by Figure \ref{fig_dir_bvp4c}, 
%which displays the results obtained by the \textsc{Matlab} routine 
%\emph{bvp4c} when $(3 \beta^2 - 2)\gamma - 4 =1$.
%\begin{figure}
% \centering
% \includegraphics[width=0.9\textwidth,height=0.5\textwidth]{fig_six.pdf}
% \caption{{\small 
% Plots of the displacement and of its first derivative through the slab thickness in 
% the case of transverse isotropy (one family of fibers). 
% Numerical solutions for the Dirichlet BVP obtained by the \textsc{Matlab} code 
% \emph{bvp4c} for $\beta=1.0$, $\gamma=5.0$, $\alpha=10.0$.}}
% \label{fig_dir_bvp4c}
%\end{figure}
Note that the non-monotonous behavior does not necessarily occur within 
the slab thickness and that some parameter values allow a monotonous variation 
of $f'$, devoid of jumps (such is for instance the case when $g_1 \ge 1$).
We focus on those parameter values which do give a jump inside the slab.

%In the case of Dirichlet b. c. $f(0)=f(1)=0$, the problem is more difficult since the integration 
%constant $C_1$ cannot be obtained explicitly \textit{a priori} as for the mixed b.c's. 
%This means that we cannot locate exactly the singular point $\hat \eta$. 
%To solve the problem by classical numerical techniques for BVP can yield unfeasible solutions  and
%also in this case commercial solvers present a warning about failure in the numerical convergence. 
%For example, in Figure \ref{fig_dir_bvp4c} we show the solution obtained again by means of the 
%Matlab function \emph{bvp4c}.

The main difficulty in solving the Dirichlet BVP is that we do not have an 
analytical access to the value of the integration constant $\eta_0$ in \eqref{g}.
We tackle the Dirichlet BVP by a \emph{shooting method}, combined with the 
\emph{bisection method}, in the following manner.
 
We take $\eta_0^{(0)}$ (say) as an initial guess for $\eta_0$.
Then let $K_0^{(0)} \equiv f'(0)$;
it is the real root to the cubic $g(K_0^{(0)}) = - \alpha \eta_0^{(0)}$.
It is now possible to reformulate the BVP as an Initial Value Problem (IVP),
which we solve numerically on two subintervals of $[0,1]$. 
That process is detailled later, in the simpler case of the mixed BVP.
It gives $\eta_S^{(0)}$, the thickness where the jump takes place,
and also $f(\eta)$ numerically.
Finally we compute $f(1)$ and measure how different it is from the 
second boundary condition $f(1)=0$: 
if $|f(1)| \le \;$tol is not satisfied,
for a prescribed numerical tolerance ``tol'',
then we adjust the approximate value of $\eta_0$ from 
$\eta_0^{(0)}$ to $\eta_0^{(1)}$ and so on, from 
$\eta_0^{(k-1)}$ to $\eta_0^{(k)}$,  
until  the criterion of convergence is reached, following the indications 
given by the bissection method. 
In the process we also get access to $\eta_S^{(k)}$, a numerical approximation of 
the singularity point $\eta_S$.

\begin{figure}
 \centering
\includegraphics[width=0.9\textwidth,height=0.45\textwidth]{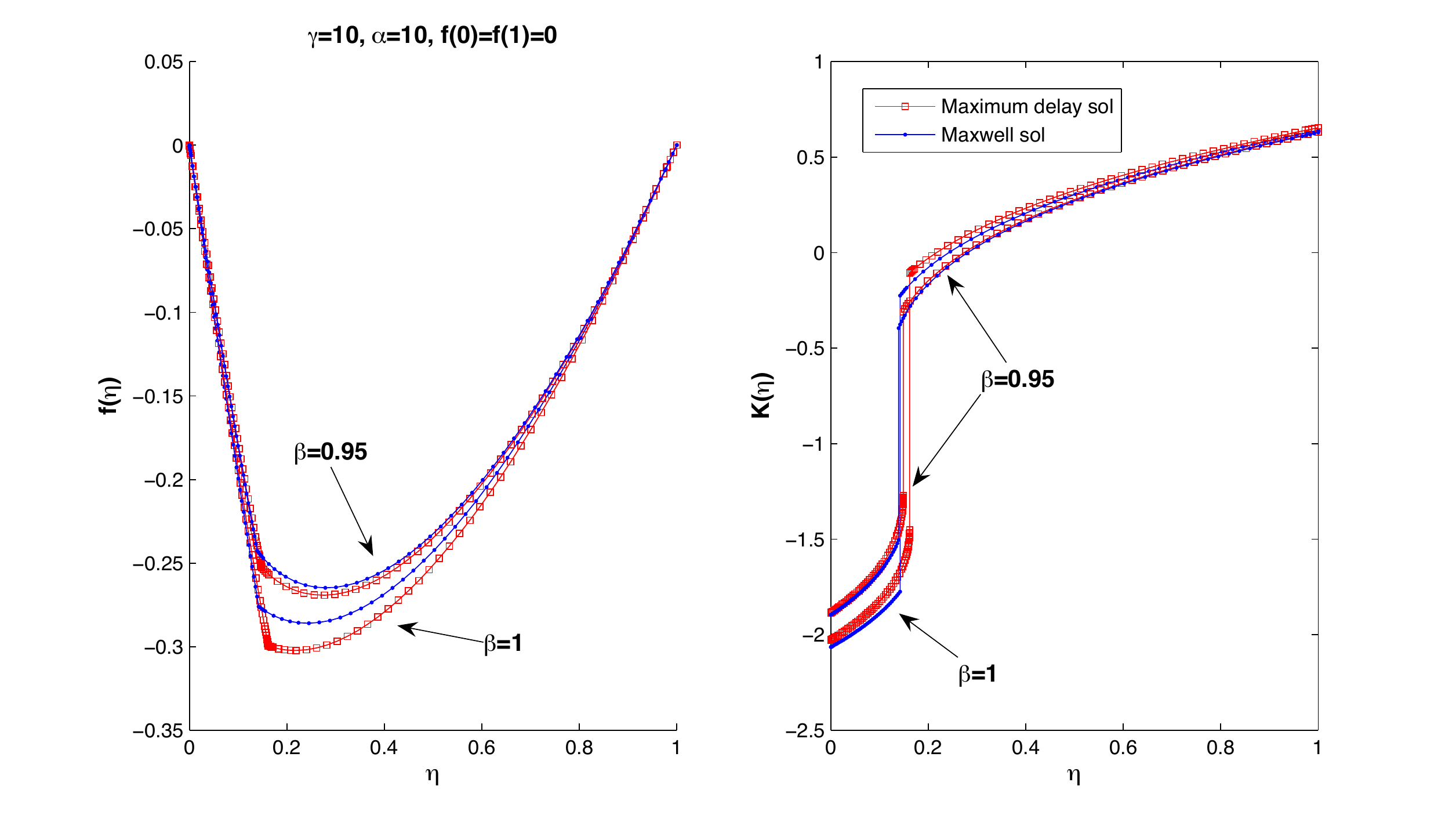}
 \caption{{\small Maxwell rule solutions and Maximum delay solutions for 
 $\beta=1.0$ (transverse isotropy) and for $\beta= 0.95$ (orthotropy), 
 with zero Dirichlet boundary conditions.
 Here $\gamma = 10.0$, $\alpha=10.0$. 
  %An iterative algorithm is used in the shooting method to identify
  %the following values of the integration constants. 
  %When $\beta=1.0$, we find $\eta_0 = 0.2423$ for the Maxwell rule solution and 
  %$\eta_0 = 0.21725$ for the Maximum delay solution; 
  %when $\beta=1.1$, we find $\eta_0 = 0.13818$ for the Maxwell rule solution 
  %and $\eta_0 = 0.05014$ for the Maximum delay solution.
  }}
 \label{fig_dir}
\end{figure}

In Figure \ref{fig_dir}, we report the numerical solutions for 
$\gamma=10.0$, $\alpha=10.0$, and in turn, 
$\beta=1.0$ and $\beta=0.95$, with tol=$1e$-6 as the tolerance 
for the stopping criterium in the bissection method. 
The values identified by the bisection method for the integration constants are as follows.
When $\beta=1.0$ (transverse isotropy), we find $\eta_0 = 0.2423$ for the Maxwell rule solution
and  $\eta_0 = 0.21725$  for the Maximum delay solution; 
when $\beta=0.95$ (orthotropy), we find  $\eta_0 = 0.13818$ for the Maxwell rule solution and 
$\eta_0 = 0.05014$ for the Maximum delay solution.
It is worth noting that the two kinds of solutions not only jump at different singular points, 
but also present different slopes, before and after the singular points.
Moreover, we checked that the solutions obtained for $\beta=1.0$ (transverse isotropy) 
are consistent with those obtained by Merodio \emph{et al.} \cite{MeSS07}, using 
a different numerical method, based on a quadrature approach.

%***************************************************

%\subsection{The mixed BVP and its numerical solution}

%****************************************************

We now consider the mixed BVP, Case \emph{(ii)},
\be \label{normalformstand2}
\dfrac{\text{d}^2 f}{\text{d} \eta^2} = \dfrac{\alpha}{D(f', \Phi)},
\qquad
f(0)=0, \qquad f'(1)=K_1,
\en
which turns out to be simpler to analyze and to  solve numerically
than the Dirichlet BVP.

The main features uncovered in the previous analysis still apply. 
Hence the uniqueness of the solution is not  guaranteed for all parameter values, 
because the energy can have two minima and is in general not a convex function,  
leading to a jump in the derivative of the displacement. 
The analysis for the mixed boundary conditions is almost identical to that of the Dirichlet
boundary conditions, with the difference that it is now possible
to identify \textit{a priori} the location of the singularity. 

In order to jump following the absolute minima of the energy, again we consider in turn the 
\emph{Maxwell rule convention} of equal area, and the \emph{Maximum delay convention}, 
because commercial solvers also fail here.
We track these solutions by 
transforming the mixed BVP into a second-order initial value problem (IVP), as follows.

Starting from the first integral \eqref{g}-\eqref{cubic} 
of equation \eqref{equationstandard2},
%$W_K(K) =K + \gamma K \sin^2\Phi [2\cos^2\Phi + 3 \beta K \sin \Phi \cos \Phi + K^2\sin^2\Phi]$, it is
%possible to calculate  the integration constant $C_1$ by the given 
we find from the second boundary condition $f'(1) = K_1$ that 
\begin{equation} \label{mix2}
\eta_0 = 1 - g(K_1)/\alpha.
\end{equation}
Then let $K_0 \equiv f'(0)$; it is the real root to the cubic 
\begin{equation} \label{mix3}
%K_0 + \gamma K_0 \sin^2\Phi [2\cos^2\Phi + 3 \beta K_0 \, \sin \Phi \cos \Phi + K_0^2\sin^2\Phi]= C_1,
 g(K_0) = -\alpha \eta_0 = g(K_1) - \alpha.
\end{equation}
Now we can reformulate the BVP as an IVP, which we solve numerically in two steps.
First on the subinterval $[0, \eta_S]$, with initial conditions: 
$f(0)=0$, $f'(0) = K_0$; 
we call $f_S$ and $K_S$ the computed values of $f$ and $f'$ at 
$\eta = \eta_S$, the slab thickness where the jump takes place.
Next we solve numerically the second part of the IVP, this time on the 
subinterval $[\eta_S, 1]$, with initial values: $f(\eta_S) = f_S$, $f'(\eta_S) = K_S$.
To compute the value of $\eta_S$, the singularity thickness, we proceed as follows.

In the case of the \emph{Maxwell rule convention} of equal area, 
the singularity occurs at the inflection point of the function $g$. 
Solving $g''(K_S) = 0$ gives $K_S = - \beta$ and then,  
$g(K_S) = \beta[\gamma(\beta^2-1)/2 -1]$.
Then $\eta_S$ is found by solving the equation 
\be \label{eta_S}
g(K_S) = g(K_1) + \alpha(\eta_S - 1).
\en

In the case of the \emph{Maximum delay convention}, 
the singularity occurs at the local maximum of the function $g$. 
Hence $K_S = x_1$ given by \eqref{x_1};
then $g(K_S) = g_1$ given by \eqref{g_1}, and $\eta_S$ is 
found from \eqref{eta_S} for $\eta_S$.

\begin{figure}
 \centering
\includegraphics[width=0.9\textwidth,height=0.45\textwidth]{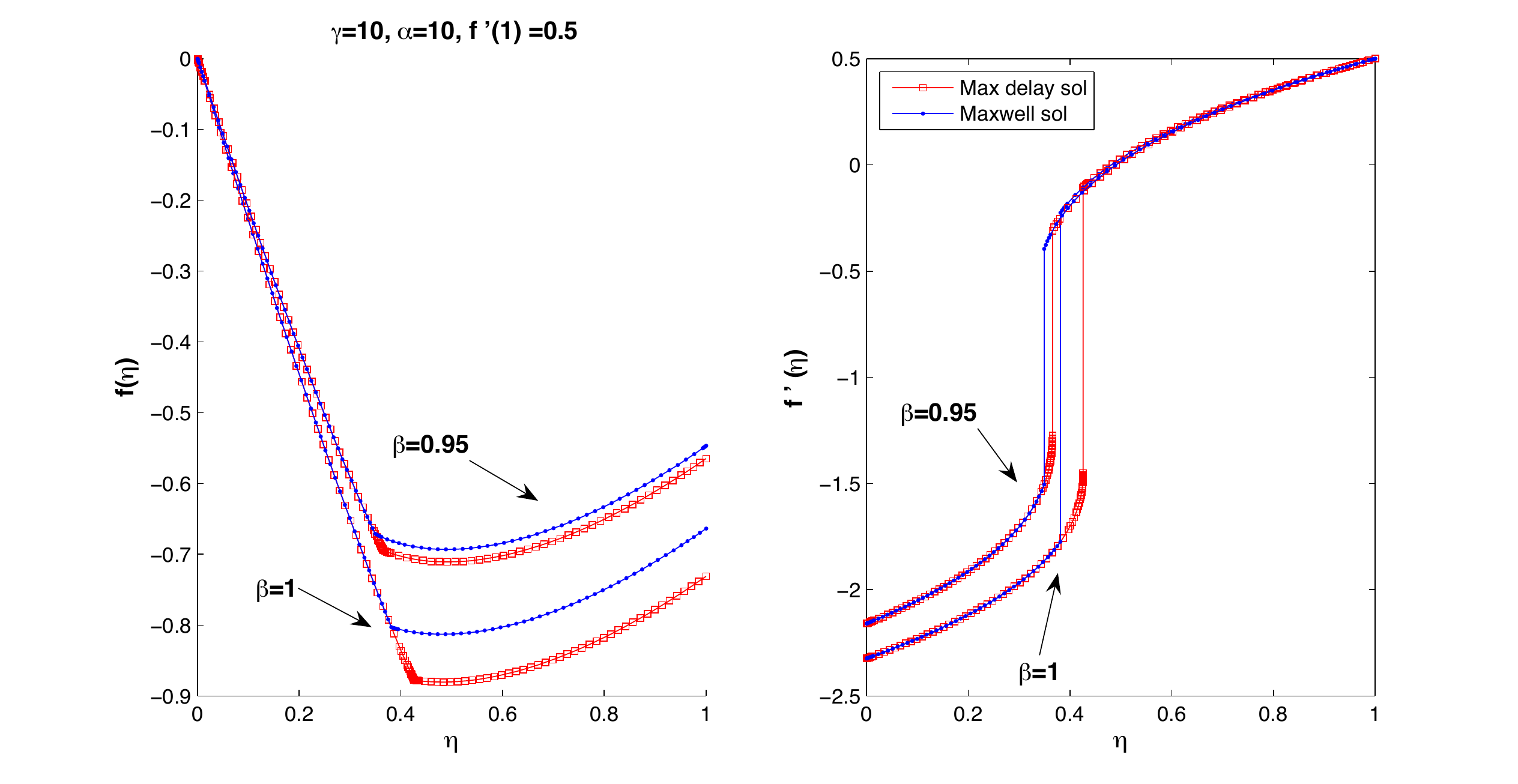}
 \caption{{\small Plots of the displacement and of its first derivative through the slab thickness in 
 the case where the two families of fibers are orthogonal. 
 Numerical solutions  for the mixed BVP obtained by following the Maxwell rule convention 
 (round dots plots) and the Maximum delay convention (square dots plots),
 for $\beta=1.0$ (transverse isotropy) and $\beta = 0.95$ (orthotropy), 
 and $\gamma=10.0$, $\alpha=10.0$, $K_1 = f'(1) = 0.5$.}}
\label{fig_mix}
\end{figure}

Figure \ref{fig_mix} shows the numerical solutions obtained with 
this numerical technique for the values $\gamma=10.0$, $\alpha=10.0$, $K_1=0.5$,  
and in turn, $\beta=1.0$ (transverse isotropy) and $\beta= 0.95$ (orthotropy). 
In the figure on the left, we report the numerical approximation for $f(\eta)$ and in the figure on 
the right, the approximations for the amount of shear $f'(\eta)$, clearly 
showing that the jumps of the
derivatives occur at different singular points.
For that example, we find $\eta_0 = 0.48125$ when $\beta=1.0$ and 
$\eta_0 = 0.44375$ when $\beta= 0.95$.

%***************************************************

\subsection{Non-orthogonal fibers: $\Phi \ne \pi/4$}

%****************************************************

%If we set
%$$
%\gamma^{*}=\frac{4}{(3 \beta^2-2)\sin^2(2\Phi)}
%$$
%and we require $\gamma=\gamma^{*}$ the real root of the cubic equation
%$K M(K,\Phi)=\delta(X_3-C_1)$ may be found in a compact
%form and therefore the exact solution of the BVP may be written down as
%\begin{align}\label{eqcompleta} \notag
%f(X_3) & =  \frac{3\sin^{2/3}(2\Phi)}{16(3\beta^2-2)\delta\sin^4(\Phi)}
%\left\{ \left[ 2\delta \sin^2(\Phi)(3 \beta^2-2) (X_3-C_1)
% +\beta^3 \sin(2 \Phi) \right]^{4/3} \right.
%\\ \notag
% & +  \left. \left[ \beta^3 \sin(2 \Phi)
% - 2 \delta\sin^2(\Phi)(3 \beta^2-2) C_1   \right]^{4/3} \right\}\\
% & - \beta \frac{\sin(\Phi)}{2\sin^2(\Phi)}X_3,
%\end{align}
%where the integration constant $C_1$ must be determined
%from the algebraic equation $f(1)=0$.
%We point that this is an important special case because
%it gives the critical value
%in correspondence where the onset of singular behaviour may appear.

Extending the results and techniques developed at $\Phi = \pi/4$ to the case $\Phi \ne \pi/4$ 
(non-orthogonal fibers) poses no particular problem.
Rather than detailing the process, we refer the reader to the paper by Merodio \emph{et al.}, 
\cite{MeSS07} where the extension is done in the case $\beta = 1$ (transverse isotropy).

%%%%%%%%%%%%%%%%%%%%%%%%%%%%%%%%%%%%%%%%%

\section{Orthotropic biomechanical model}

%%%%%%%%%%%%%%%%%%%%%%%%%%%%%%%%%%%%%%%%

We now investigate briefly whether the analysis conducted for
the standard reinforcing model can be
extended to a strain energy density often
encountered in the biomechanics literature,
namely the model proposed by Holzapfel \emph{et al.} \cite{HoGO00}
to describe the behavior of an orthotropic
artery, and widely used since, for instance to model porcine aortic tissue,
passive basilar artery, cornea, etc.
We present it in the form
\begin{multline} \label{Holzapfel1}
W=\dfrac{\mu}{2}(I_{1}-3) +
\dfrac{\mu E_1}{2k_1}\left\{\exp\left[k_1(I_{4}-1)^2\right]-1\right\}
\\
+ \dfrac{\mu E_2}{2 k_2}\left\{\exp\left[k_2(I_{6}-1)^2\right]-1\right\},
\end{multline}
where $k_1$, $k_2$ are dimensionless constants.
Equation (\ref{equation}) is then rewritten as
\begin{multline}
\frac{d}{d\eta}
 \left\{f' +E_1(I_4-1)\exp\left[k_1(I_{4}-1)^2\right] m_1 m_3 \right.
\\
 \left. +  E_2(I_6-1)\exp\left[k_2(I_{6}-1)^2\right]n_1 n_3\right\}
 = C_0 L / \mu.
\end{multline}

The BVP can be put in the form \eqref{normalformstand}, where now
\begin{multline}
D(f', \Phi)=1+\sin\Phi
\left\{E_1\Gamma_1(f', \Phi)\exp\left[k_1(I_{4}-1)^2\right]\right.
\\
\left. +E_2\Gamma_2(f', \Phi)\exp\left[k_2(I_{6}-1)^2\right]\right\},
\end{multline}
and the functions $\Gamma_1$ and $\Gamma_2$ are defined as
\begin{multline}
\Gamma_1=2k_1 f'^2 \left(f' \sin\Phi + \cos\Phi \right)
  \left(f'\sin^2\Phi + \sin 2\Phi \right)^2
   \left(2f'\sin^2\Phi + \sin 2\Phi \right) \\
+ 3 (f')^2\sin^3 \Phi + 3 f'\sin\Phi \sin2\Phi +\cos\Phi \sin2\Phi,
\end{multline}
and
\begin{multline}
\Gamma_2 = 2k_2 f'^2 \left(f' \sin\Phi - \cos\Phi \right)
 \left(f'\sin^2\Phi-\sin2\Phi\right)^2
  \left(2f'\sin^2 \Phi - \sin 2\Phi\right)
\\
- 3 (f')^2\sin^3 \Phi +3  f' \sin\Phi \sin2\Phi + \cos\Phi \sin2\Phi.
\end{multline}

To simplify the algebra we restrict our discussion
to the special case where the families of fibers
are at right angle, $\Phi = \pi/4$.
Our objective is to find out if there exist special values
$f' = (f')^*$ say, such that $D((f')^*, \pi/4)=0$.
Now $\Gamma_1$ and $\Gamma_2$ reduce to
\begin{align}
& \Gamma_1(f', \pi/4)=\frac{\sqrt{2}}{4}
\left\{f'^2(f'+1)^2(f'+2)^2k_1+(3f'^2+6f'+2) \right\},
\notag \\
& \Gamma_2(f', \pi/4)=\frac{\sqrt{2}}{4}
\left\{f'^2(f'-1)^2(f'-2)^2k_2+(3f'^2-6f'+2) \right\}.
\end{align}

In the biomechanical applications of the
model \eqref{Holzapfel1}, it is often assumed that the
two families of fibers are mechanically equivalent,
so that $E_1=E_2$ and $k_1=k_2$.
In that case, some long but simple computations show
that $f'\equiv0$ is a minimum for the function $D$.
Because $D(0, \pi/4)=1+E_1 \neq 0$, we conclude that
singularities may not develop
(This result may be extended to any angle $\Phi$ quite easily).

\begin{figure}
 \centering
\includegraphics[width=0.9\textwidth,height=0.45\textwidth]{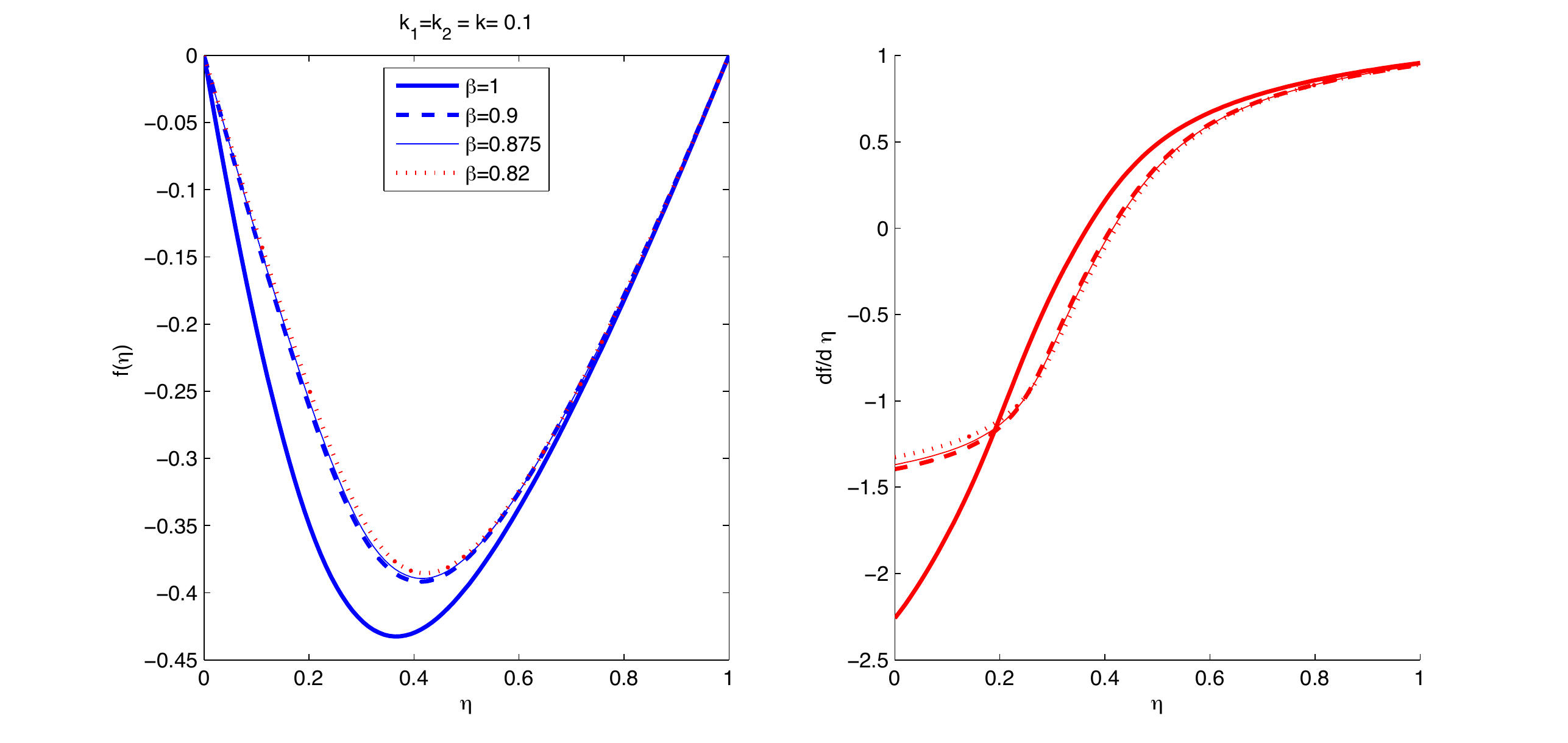}
 \caption{{\small Numerical solution to the zero Dirichlet BVP for the aartery model.
 Here $k_1 = k_2 = 0.1$ and $\beta = 1.0$ (transverse isotropy),
 $0.9$, $0.875$, $0.82$ (orthotropy).
 Other parameters: $E_1 + E_2 = 10.0$, $C_0 L / \mu = 10.0$.}}
 \label{fig_exp}
\end{figure}

When $E_1 \neq E_2$, things are more complex.
For instance, consider the values of $D(f', \pi/4)$ when
$f' = -1,0,1$ in turn:
\begin{align}
&D(-1, \pi/4)= 1 - E_1\ee^{k_1/4} + (36k_2+11) (E_2/4)\ee^{3k_2/2} , \nonumber
\\
&D(0, \pi/4)= 1 + (E_1+E_2)/2,
\nonumber
\\
& D(1, \pi/4) = 1 + (36k_1 + 11)(E_1/4) \ee^{3 k_2/2} - E_2 \ee^{k_2/4}.
\end{align}
Therefore, because $D(0, \pi/4)>0$, it is sufficient to choose
\be
\exp(k_{1}/4)E_1>1+\frac{36k_2+11}{4} \exp(9k_2/4)E_2
\en
to obtain the existence of at least one $(f')^*$ such that
$D((f')^*,\pi/4)=0$.
This inequality suggests that singularities occur only for huge differences between the fiber stiffnesses, 
and are unlikely to be observed at all for realistic values of the parameters. 
Take for example the case where $k_1 = k_2 = k$ (say). 
Then $\beta$ defined in \eqref{alpha_beta} gives a measure of the orthotropy:
when $\beta=1$, the solid is reinforced with one family of parallel fibers, 
and when $\beta<1$, there are two families of parallel fibers at play. 
To generate the graphs in Figure \ref{fig_exp}, we take $k = 0.1$ and $\beta = 1.0$, $0.9$, $0.875$, 
and $0.82$ in turn. 
At $\beta=1.0$, the shear variations are pronounced but regular. 
As soon as $\beta<1$ (two families of fibers), the shear variations are quickly smoothed down,
highlighting the stabilizing effect of orthotropy.

We now evoke some possible applications of our results  to biomechanics.
Indeed, we know that arterial tissue adapts to physiological and
pathological stimuli though rearrangement of the microstructure.
Arterial remodeling is induced by chronically altered mechanical forces;
if for some pathological reason, the remodeling of
the fibers introduces some disparity in the various directions in the
stiffness of the fibers, then it may happen that $E_1 \neq E_2$ and
that some ``dangerous'' mechanical behavior develops.
However, from a mathematical point of view the solutions of the BVPs suggest that the artery 
model is much more stable than the standard reinforcing model, due to the presence of 
exponential terms in the determining equations.

%%%%%%%%%%%%%%%%%%%%%%%%%%%%%%%

\section{Concluding remarks}

%%%%%%%%%%%%%%%%%%%%%%%%%%%%%%%

We extended the results of Merodio \emph{et al.} \cite{MeSS07}
from transverse isotropy to orthotropy.
The most important finding is that  orthotropic
materials may develop singular solutions
only if there is a significant difference between
the mechanical stiffnesses of the two families of fibers.
We quantified this result rigorously for
the standard reinforcing model,
where a necessary condition for the formation
of singular solutions is that
$\beta^2>2/3$, which means that one family of fibers must
be at least 9.9 times stiffer than the other family.

When we consider the arterial strain-energy density \eqref{Holzapfel1},
analytical results are no longer possible,
but the methodology used to study the standard
reinforcing material is still applicable.
In this case a huge difference between $E_1$ and $E_2$ is necessary
to possibly introduce a singularity.
However, if the fibers are mechanically equivalent, as is usual for biological soft tissues,
then singular solutions are avoided altogether.
Therefore biological networks, such as the collageneous structure of
arterial walls,
are the \textit{right} structure to prevent
the formation of the singularities
described here.

From a theoretical point of view, our results demonstrate the complexity of
finite anisotropic elasticity and deliver some exact solutions,
which are scarce in the literature on finite inhomogeneous deformations
of orthotropic materials.

It is important to note that we have barely scratched the
surface of the collection of problems associated with the
rectilinear shear of solids reinforced by two families of
parallel fibers.
\emph{Primo}, we relied on strong ---and perhaps, reductive---
constitutive assumptions, namely that the strain energy density
can be split into the sum of an isotropic part
and an anisotropic part, and that this latter part is also
the sum of two parts, each depending on only one anisotropic
invariant.
Although there is now a good body of experimental data supporting the
adequacy of the standard reinforcing model \eqref{standard} and of the
biomechanics arterial model \eqref{Holzapfel1}, the importance
or insignificance of other constitutive arguments must also be
evaluated, such as the role played by other invariants \cite{MeOg06}
or by the angular distribution of fiber directions
\cite{MNAH07, GaOH06, PaHo08}.
\emph{Secondo}, we limited our study to a shear occurring along the bissectrix of the two
families of parallel fibers, and did not study the influence of
other orientations.
Intuitively, it is expected that this is the direction where the coupled
reinforcing effect of the fibers is at its strongest.
Nevertheless we were able to show that if one family of fibers
is much stiffer than the other for the standard reinforcing model, then singularities might
develop in the thickness of the clamped slab, in the form of
discontinuities in the shear or in the strain gradient.

%++++++++++++++++++++++++++

%++++++++++++++++++++++++++++++++++++++++++++++++++++++++++++++++


\begin{thebibliography}{99}
%++++++++++++++++++++++++++

{\small

\bibitem{Antm95}
S.S. Antman,
Nonlinear Problems of Elasticity,
Springer Verlag, New York, 1995.

\bibitem{Adki55}
J.E. Adkins,
%Some general results in the theory of large elastic
%deformations.
Proc. Roy. Soc. London A  231 (1955) 75-90.
%
\bibitem{ErRi54}
J.L. Ericksen, R.S. Rivlin,
%Large elastic deformations of
%homogeneous anisotropic materials,
J. Rational Mech. Analysis 3 (1954) 281-301.

\bibitem{Lekh68}
S.G. Lekhnitskii,
Anisotropic Plates,
Gordon \& Breach, New York (1968).

\bibitem{AnNe87}
S.S. Antman, P.V. Negron Marrero,
%The remarkable nature of radially symmetric states
%of aelotropic nonlinearly elastic bodies.
J. Elasticity 18 (1987) 131-164.

\bibitem{KMOP07}
F. Kassianidis, J. Merodio, R.W. Ogden, T.J. Pence,
%Azimuthal Shear of a Transversely Isotropic Elastic Solid
Math. Mech. Solids (to appear).

\bibitem{MeSS07}
J. Merodio, G. Saccomandi, I. Sgura,
%The rectilinear
%shear of fiber-reinforced incompressible non-linearly elastic
%solids.
Int. J. Non-Linear Mech. 42 (2007) 342-354.

\bibitem{FoRo01}
R. Fosdick, G. Royer-Carfagnig,
%The constraint of local
%injectivity in linear elasticity theory,
Proc. Roy. Soc. London A 457 (2001) 2167-2187.

\bibitem{HoGO00}
G.A. Holzapfel, T.C. Gasser, R.W. Ogden,
J. Elasticity 61 (2000), 1-48.

\bibitem{Spen72}
A.J.M. Spencer,
Deformations of fiber-Reinforced Materials.
University Press, Oxford (1972).

\bibitem{Boel87}
J.P. Boehler,
Applications of tensor functions in solid mechanics,
in:
CISM Lecture Notes 292, Springer Verlag, New York (1987).

%Danescu, A., 1991. Bifurcation in the traction problem for a transversly
%isotropic material, Math. Proc. Camb. Phil. Soc. 110, 385--394.


%Holzapfel, G.A.2008. Collagen in Arterial Walls:
%Biomechanical Aspects. In: P. Fratzl (ed.),
%"Collagen. Structure and Mechanics", Chapter 11, Springer-Verlag, Heidelberg, 285--324.

%Humphrey, J. D., 2002. Cardiovascular Solid Mechanics, Cells, tissues and
%Organs, Springer, New York.

%Qiu, G. Y. and Pence, T. J., 1997. Remarks on the behavior of simple
%directionally reinforced incompressible nonlinearly elastic spheres, J.
%Elasticity 49, 1--30.

%Triantafyllidis N. and Abeyaratne, R., 1983. Instability of a finitely
%deformed fiber-reinforced elastic material, J. Appl. Mech. 50, 149--156.

%Tsai H. and Fan,\ X., 1999. Anti-plane shear deformations in
%compressible transversely isotropic materials, \ J. of Elasticity
%54, 73-88.

\bibitem{MeOg06}
J. Merodio, R.W. Ogden,
%The influence of the invariant I8  on the
%stress-deformation and ellipticity characteristics of doubly
%fiber-reinforced non-linearly elastic solids
Int. J. Non-Linear Mech. 41 (2006) 556-563.

\bibitem{MNAH07}
A.S. Milani, J.A. Nemes, R.C. Abeyaratne, G.A. Holzapfel,
%A method for the approximation of non-uniform fiber misalignment
%in textile composites using picture frame test.
Composites A: Appl. Sc. Manufact. 38 (2007) 1493-1501.

\bibitem{GaOH06}
T.C. Gasser, R.W. Ogden, G.A. Holzapfel,
%Hyperelastic modelling of arterial layers with distributed collagen
%fiber orientations (with ), Journal of the
Roy. Soc. Interface 3 (2006) 15-35.

\bibitem{PaHo08}
A. Pandolfi, G.A. Holzapfel,
%Three-dimensional modeling and computational analysis of the human
%cornea considering distributed collagen fibril orientations. ASME
J. Biomech. Eng. (in press).

}


\end{thebibliography}
\end{document}